\documentclass[fleqn,usenatbib]{rasti}

\usepackage{newtxtext,newtxmath}
\usepackage[T1]{fontenc}

\DeclareRobustCommand{\VAN}[3]{#2}
\let\VANthebibliography\thebibliography
\def\thebibliography{\DeclareRobustCommand{\VAN}[3]{##3}\VANthebibliography}

\usepackage{graphicx}	% Including figure files
\usepackage{amsmath}	% Advanced maths commands
\usepackage{orcidlink}
\usepackage{physics}

\newcommand{\Mmin}{M_{\rm min}}
\newcommand{\cinemas}{\textsc{cinemas}}
\newcommand{\rebound}{\textsc{rebound}}
\newcommand{\spock}{\textsc{spock}}

\title[The true masses of radial-velocity exoplanets]{The true masses of radial-velocity exoplanets constrained by stability}

\author[X.\ Byrne et al.]{
Xander Byrne
\orcidlink{0000-0001-9488-238X}$^{1}$\thanks{E-mail: xbyrne@ast.cam.ac.uk} and
Amy Bonsor
\orcidlink{0000-0002-8070-1901}$^{1}$
\\
$^{1}$Institute of Astronomy, University of Cambridge, Madingley Road, Cambridge CB3 0HA, UK\\
}

\date{Accepted XXX. Received YYY; in original form ZZZ}

\pubyear{\the\year{}}

\begin{document}
\label{firstpage}
\pagerange{\pageref{firstpage}--\pageref{lastpage}}
\maketitle

\begin{abstract}
Exoplanets discovered with the radial velocity method suffer from a mass-inclination degeneracy: only the \textit{minimum} mass $\Mmin=M\sin i$ is measured.
An exoplanet with an Earth-sized minimum mass could actually therefore be a sub-Neptune, or even a gas giant.
However, for compact multi-planet systems, the inclination angle cannot be too low, as this would imply masses so large that the system would be dynamically unstable on time-scales shorter than the age of the system.
We present \cinemas, a Bayesian framework to constrain the inclinations -- and hence true masses -- of radial-velocity exoplanets in flat multi-planet systems, by accounting for dynamical stability.
Applying this framework to five compact exoplanetary systems, we cut upper limits on the true masses of their planets by as much as 45\%.
As two examples, the masses of the four Barnard's Star planets are constrained to all be below $0.6M_{\earth}$ to at least 95\% confidence; and the three gas giants orbiting HD~184010 are found to be intermediate in mass between Saturn and Jupiter.
With only $\sim1\%$ of planetary systems being inclined enough to transit, methods of studying non-transiting systems are vital for constraining the properties of nearby exoplanets.
\end{abstract}

% Include between one and six keywords.

%%%%%%%%%%%%%%%%%%%%%%%%%%%%%%%%%%%%%%%%%%%%%%%%%%

%%%%%%%%%%%%%%%%% BODY OF PAPER %%%%%%%%%%%%%%%%%%

\section{Introduction}

The radial velocity (RV) technique has been one of the most important methods for finding planets beyond the Solar System, responsible for the discovery of the first \citep{wolszczan92, mayor95} and nearest \citep{angladaescude16, gonzalezhernandez24, diaz19} exoplanets.
By measuring shifts in stellar spectral lines due to the gravitational influence of planets, the RV technique yields orbital periods, eccentricities and minimum masses for each detected planet.

Although the transit method \citep{charbonneau00} has been much more prolific in exoplanet discovery \citep[e.g.,][]{batalha13, guerrero21}, this is primarily due to a larger search volume.
The method relies on photometric rather than spectroscopic measurements (as in the RV technique), which has been exploited by large survey missions such as CoRoT \citep{corot}, \textit{Kepler} \citep{kepler}, and TESS \citep{tess}.
The probability that a given exoplanet transits its star is geometrically very low, requiring its orbit to be almost edge-on.
This restriction is particularly important in the context of finding exoplanets around nearby stars.
The nearest known transiting exoplanets are $6.5~{\rm pc}$ away (HD~219134~b and c; \citealt{gillon17}), but there are over 100 known stars \citep{gaiadr3} and 33 planetary systems of 67 known planets \citep{christiansen25} that are closer to us.
Of these non-transiting planets, almost all were discovered using the RV technique.

The RV technique is capable of discovering non-transiting exoplanets, but the inferred masses are degenerate with the system's inclination\footnote{
    Throughout this work, \textit{inclination} refers to the angle of the plane of the planetary system to the plane of the sky.
    Inclinations of different planets in the same system relative to each other are referred to as \textit{mutual inclinations}.
} angle $i$.
The signal size for an exoplanet of mass $M$ is proportional to $M\sin i$, because only the line-of-sight motion of the star is measured.
Thus only a \textit{minimum mass} $\Mmin\equiv M\sin i$ can be deduced, in the absence of additional data.

\begin{figure}
\includegraphics[width=\columnwidth]{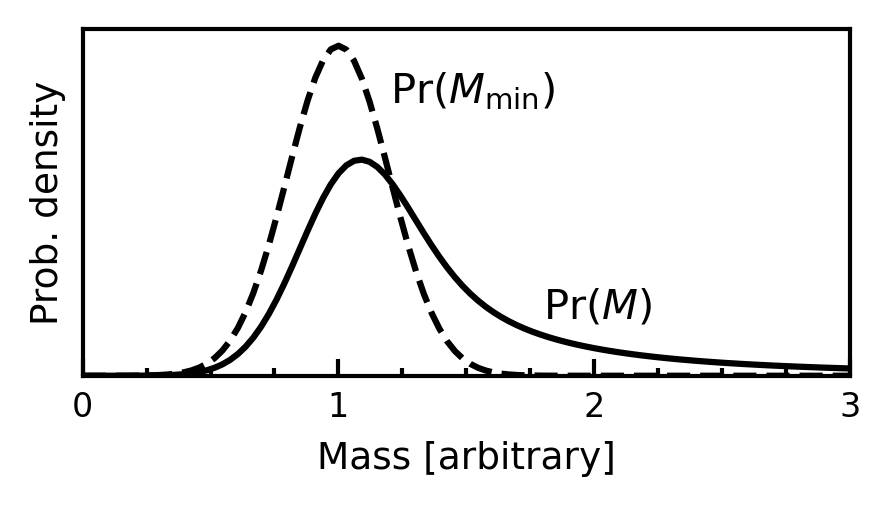}
\caption{
    The relationship between a probability distribution of the minimum mass $\Mmin$, and the corresponding distribution for the true mass $M=\Mmin/\sin i$, assuming an isotropic inclination distribution $\Pr(i)=\sin i$.
    The minimum mass distribution (perhaps as fitted to RV data) is shown as a normal distribution with mean $1.0$ and standard deviation $0.2$.
    The resulting distribution for the true mass $M$ is skewed towards larger values, has a greater expectation value by a factor $\pi/2\approx 1.6$, and has formally infinite variance.
}
\label{fig:truevmin}
\end{figure}

The usually-unknown factor of $\sin i$ between the minimum and true masses can drastically change the interpretation of the planet.
For a given probability distribution on the minimum mass (as obtained, for example, from analysis of RV data), the corresponding distribution on the \textit{true} mass has a greater mean by a factor of $\pi/2$, and a standard deviation that is formally infinite (Fig.~\ref{fig:truevmin}; see also Appendix~\ref{app:truevmin}).
For an RV-detected planet, there is a 13\% probability that its mass is more than double its minimum mass, and a 2\% chance that it is more than five times larger.
For some RV-detected companions, the subsequent determination of very low inclinations has shown them to even be above the deuterium-burning limit, and therefore to not even be planets at all \citep{kiefer20}.
This lack of knowledge of such a fundamental parameter as an exoplanet's mass has significant consequences for its inferred characteristics, including its bulk density and interior structure \citep[e.g.][]{rogers10, dorn15, rodriguezmartinez23}, thermal evolution \citep{stamenkovic12, stixrude14}, water storage capacity \citep{guimond23, byrne26}, and habitability \citep[e.g.][]{schulzemakuch11, unterborn16}.
Beyond individual planets, such uncertainties also propagate out to our understanding of the demographics of the overall exoplanet population \citep[e.g.][]{baumeister25}.

Multi-planet systems can be dynamically delicate.
Only a small region of the parameter space of orbital configurations corresponds to stable systems, partly due to the large number of mean-motion resonances that are possible for $>2$ planets \citep{wisdom80, chambers96, duncan97, quillen11}.
Under the `temporal Copernican principle' \citep[e.g.][]{gott93}, we may assume that systems we observe at the present day are dynamically stable (unless, perhaps, they are exceedingly young).
The condition of dynamical stability has been used extensively to constrain the orbital parameters of multi-planet systems \citep[e.g.][]{gozdziewski03, marshall10, fabrycky12, kosiarek21, mayo23, chachan25}, including planets' masses \citep[e.g.][]{steffen13, quarles17, fukui21, sikora23}.
This condition has also provided constraints on additional undetected planets in known systems \citep[e.g.][]{pearson19, murgas23, obertas23, pozuelos23, schwarz26}.

The delicateness of compact multi-planet systems can be exploited to constrain the orbital inclination, and hence true planetary masses, in radial-velocity systems.
The inclination cannot be too low: if it were, this would imply that the true masses $M=\Mmin/\sin i$ would be so large that the system would be dynamically unstable on short ($<{\rm system\ age}$) time-scales.

Orbital inclinations are rarely investigated in stability analyses.
Usually stability analyses are only attempted if the inclination is known, whether because they are transiting (and hence $i\approx90^\circ$), or because they are in disks \citep{tamayo15}, or because absolute astrometry is available \citep{giovinazzi26}.
However, even for RV-discovered systems with unknown inclination, stability investigations often ignore this parameter and assume $i=90^\circ$, even when the system is not transiting \citep[e.g.][]{rosenthal19, stalport22, dreizler24}.

An example of a system in which stability \textit{has} been used to constrain inclination is HR~8799: in \citet{wang18}, the inclination is fit to astrometry from high-contrast image data, and constrained by the condition of stability, using Markov-chain Monte Carlo (MCMC) methods.
Another example is the nearby Barnard's Star planetary system:
\citet{basant25} and \citet{byrne26} perform brief scans of the inclination to obtain $i\gtrsim20^\circ$, based on the planets' minimum masses and other known orbital parameters.
A third is HD~184010, for which \citet{teng22} use a similar method to obtain $i\gtrsim14^\circ$; and for the HD~176986 system \citet{nari26} scan the inclination-eccentricity space of a planet to obtain $i\gtrsim10^\circ$.
The present work combines the data modality of \citet{basant25}, \citet{byrne26}, \citet{teng22} and \citet{nari26} (i.e., RV results), with the Bayesian approach of \citet{wang18}, to constrain the inclination of RV-discovered systems with unknown inclinations.

A significant barrier to determining the stability of orbital configurations is the computational expense of $N$-body simulations.
Although they have been used extensively to test stability \citep[e.g.,][]{gozdziewski03, fabrycky12, steffen13, pearson19, deleon21, huang25}, their great computational expense means that only a restricted parameter space can be reasonably investigated, for example assuming zero eccentricity.
This issue has been significantly mitigated by the development of \spock{} \citep{tamayo20, thadhani25}, a machine-learning-based classifier which, for a given system configuration, estimates the likelihood that the system will be dynamically stable for $10^9$ orbits (of the innermost planet).
\spock{} works by first numerically integrating a system for just $10^4$ orbits, evaluating 10 summary features, and finally using an XGBoost \citep{chen16} model (trained on $\sim100~000$ fully-integrated configurations) to return the stability likelihood for this system.
As the summary feature and XGBoost evaluations take negligible time compared to the integration, \spock{} offers an acceleration of $\approx10^5$ over full $N$-body simulations.
This enables the evaluation of stability over a much larger parameter space.

Assuming an isotropic inclination distribution, the prior probability distribution of the inclination is $\pi(i)=\sin i$.
However, low $i$ corresponds to large masses, which makes the stability of the system less likely.
This work develops the first stability-based Bayesian constraints on the inclinations of non-transiting exoplanetary systems, without the need for any additional observations.
We present \cinemas{} (\textbf{C}onstaining \textbf{IN}clinations of \textbf{E}xoplanets and their \textbf{MA}sses by \textbf{S}tability), a publicly-available\footnote{
    \url{https://github.com/xbyrne/cinemas}
} code implementing this framework.
Section~\ref{sec:methods} describes the \cinemas{} framework and the data we apply it to, and Section~\ref{sec:results} presents the results of this analysis.
We discuss these results in Section~\ref{sec:discussion} and conclude our work in Section~\ref{sec:conclusion}.

\section{Methods}
\label{sec:methods}

This section describes our approach in exploiting stability to constrain the orbital inclinations -- and hence true masses -- of exoplanets in compact non-transiting systems.
We also describe a method of constraining the masses of additional undetected planets in Section~\ref{sec:methods_undetected}.

\subsection{Bayesian framework}
\label{sec:methods_framework}

We first assume planetary systems to be planar, with no mutual inclinations\footnote{
    This assumption is discussed further in Section~\ref{sec:discussion_limitations_mutualinclinations}.
}.
For a system with a given inclination angle $i$, we have from Bayes' Theorem:
\begin{equation}
\label{eq:simplebayes}
\Pr(i|{\rm stable}) \propto \mathcal{L}({\rm stable} | i) \times \pi(i).
\end{equation}
The first term on the right-hand side of equation~(\ref{eq:simplebayes}) is the likelihood that the system is stable for a particular inclination.
For example, lower values of $i$ have lower stability likelihoods, given that this corresponds to more massive planets.
The second term is the prior; we assume an isotropic prior of $\pi(i)=\sin i$ throughout.
Their product, the posterior on the left-hand side, gives a 1D probability distribution over $i$.

The stability of a planetary system also depends on additional parameters, which have their own measurement uncertainties.
These include:
\begin{itemize}
\item \textbf{Minimum masses}.
    The minimum mass $\Mmin$ is determined from RV measurements, and will have an uncertainty associated with it.
    The `true value of the minimum mass' does impact the stability, though of course this is degenerate with inclination: if the minimum mass is lower, then lower values of the inclination will permit the same true mass, which is of course what actually determines stability.
\item \textbf{Orbital periods}.
    These are, again, directly measured in RV observations, generally with a much smaller uncertainty than the minimum masses.
\item \textbf{Stellar mass}.
    The stellar mass is not generally measured directly from RV data, but is often available (with some uncertainty) from well-calibrated spectroscopic or photometric relations.
\item \textbf{Other orbital elements}.
    Given that we restrict our analysis to flat systems for now, we limit the inclusion of additional orbital elements to the eccentricities $e$, longitudes of periastron $\varpi$, and initial true anomalies $f_0$ at some reference time.
    Without loss of generality, we need only consider the longitudes of periastron and initial true anomalies relative to that of the inner planet, respectively $\Delta\varpi$ and $\Delta f_0$.
    Eccentricities are constrained somewhat by RV measurements.
    Longitudes of periastron and true anomalies (at some epoch) can also be constrained by observations, but usually only very weakly.
\end{itemize}
The stability of a given planetary system can therefore be determined from the parameters $M_*$, $i$, $\{M_{{\rm min},p}\}$, $\{P_p\}$, $\{e_p\}$, $\{\Delta\varpi_p\}$, and $\{\Delta f_{0,p}\}$, where $p$ denotes a particular planet in the system.
Denoting the nuisance parameters (i.e.\ $M_*$, $\{M_{{\rm min},p}\}$, $\{P_p\}$, $\{e_p\}$, $\{\Delta\varpi_p\}$, and $\{\Delta f_{0,p}\}$) by $\vb{\Theta}$, we therefore complete our Bayesian framework by writing
\begin{equation}
\label{eq:fullbayes}
\Pr(i, \vb{\Theta} \ |\ {\rm stable}) \propto \mathcal{L}({\rm stable} \ |\ i, \vb{\Theta} ) \times \pi(i, \vb{\Theta}).
\end{equation}
Using $M=\Mmin/\sin i$, the posterior can be marginalised to obtain a posterior probability distribution on the true masses of each planet.

\subsection{Planetary systems studied}
\label{sec:methods_systems}

The systems studied in this work are selected using data from the NASA Exoplanet Archive (NEA; \citealt{christiansen25}).
We first select multi-planet ($N_p\geq3$) systems in which all planets were detected using the radial velocity method.
We then select systems which are \textit{compact}: following \citet{tamayo20}, we describe a system as compact if there is at least one set of three planets for which both adjacent period ratios are $<2$.
It is these systems where the stability condition can provide the strongest constraints\footnote{
`Compact' systems are also those on which the \spock{} stability classifier, which we employ in this work, is trained.
Although one could conceive of other definitions of `compactness', we use that of \citet{tamayo20} to avoid extrapolating beyond the training set of \spock.
}.
We exclude the GJ~667~C and DMPP-1 (HD~38677) systems: while purportedly satisfying all the above criteria \citep[][]{angladaescude13, staab20}, the existence of some of their reported planets has been called into question or ruled out \citep{feroz14, robertson14, standing26}, and the non-existence of these planets would disqualify their systems from the `compact' criterion.
We also exclude the HD~158259 system: while this is also a compact system, its inclination is well-constrained by the fact that the inner planet (b) transits the star, but none of the others (c--f) transit, thus restricting its inclination to between $80$ and $83^\circ$ \citep{hara20, lovos22}.

Five systems remain; important data are summarised in Table~\ref{tab:systems}, and a brief vignette of each system is presented below.

\begin{table}
\centering
\caption{
    Basic observational data for five compact multi-planet radial-velocity systems.
}
\begin{tabular}{rcrr}
\hline
Star (mass) & Planet & $P$~[d] & $M_{\rm min}$~[$M_{\earth}$] \\

\hline
HD 215152 & b & 5.76 & 1.82 \\
($0.77~M_{\sun}$) & c & 7.28 & 1.72 \\
 & d & 10.86 & 2.80 \\
 & e & 25.20 & 2.88 \\
\hline
Barnard's Star & d & 2.34 & 0.26 \\
($0.16~M_{\sun}$) & b & 3.15 & 0.30 \\
 & c & 4.12 & 0.34 \\
 & e & 6.74 & 0.19 \\
\hline
HD 184010 & b & 286.60 & 98.53 \\
($1.35~M_{\sun}$) & c & 484.30 & 95.35 \\
 & d & 836.40 & 143.02 \\
\hline
HD 28471 & b & 3.16 & 3.72 \\
($0.98~M_{\sun}$) & c & 6.12 & 5.72 \\
 & d & 11.68 & 4.91 \\
\hline
YZ Ceti & b & 2.02 & 0.70 \\
($0.14~M_{\sun}$) & c & 3.06 & 1.14 \\
 & d & 4.66 & 1.09 \\
\hline
\end{tabular}

\label{tab:systems}
\end{table}

\subsubsection{HD 215152}

All four of HD~215152's planets (b--e) have super-Earth minimum masses and periods on the order of days \citep{delisle18}.
The planets were discovered from HARPS \citep{harps} observations spanning 13 years.
Non-detections of astrometric excess noise in \textit{Gaia} DR1 data eliminate ultra-low inclinations ($i\lesssim0.2\degr$), and hence true masses greater than a few Jupiter masses \citep{kiefer21}.

\subsubsection{Barnard's Star}

Barnard's Star is the nearest single star to the Sun, an M4 dwarf at $1.8~{\rm pc}$.
It hosts a system of four sub-Earth planets (d, b, c, e in order of orbital distance), at periods between $2.34$--$6.74~{\rm d}$ \citep{gonzalezhernandez24, basant25}, discovered by combining observations from ESPRESSO \citep{espresso} and MAROON-X \citep{maroonx} between 2019 and 2023.
Notably, the outermost planet (e) has the smallest RV semi-amplitude of any confirmed exoplanet to date ($22.1~{\rm cm~s}^{-1}$; \citealt{basant25}).
The planets have minimum masses between $0.193$--$0.335~M_{\earth}$; although low inclinations would allow for these planets to be much larger, earlier stability analyses have deemed super-Earth masses improbable \citep{basant25, byrne26}.

\subsubsection{HD 184010}

HD~184010 is a markedly different system to the others studied here.
The star is an evolved K0~III star \citep{hipparcos} targeted by the Okayama Planet Search Program \citep{sato05}.
RV observations with the HIDES spectrograph \citep{hides} between 2005 and 2022 \citep{teng22} revealed three giant planets (b--d; $\Mmin$ between $0.26$--$0.46~M_J$) with periods on the order of years.
This system is unique in our study in several ways: it is the only system hosted by a post-main-sequence star, and contains by far the most massive planets, on by far the longest-period orbits.

\subsubsection{HD 28471}

HD~28471, a G5 star at $44~{\rm pc}$, hosts three super-Earths on short-period orbits, based on 19 years of HARPS observations \citep{stevenson25}.

\subsubsection{YZ Ceti}
\label{sec:methods_systems_yzceti}

Another very nearby star ($3.7~{\rm pc}$), YZ Ceti is an M4 dwarf hosting three planets with Earth-like minimum masses on short-period orbits \citep{stock20}, first identified in 2017 from 211 HARPS spectra \citep{astudillodefru17}.
Although some subsequent analyses had called the three-planet solution into question \citep{robertson18, tuomi19}, 229 additional spectra from CARMENES \citep{carmenes} were able to confirm all three planets \citep{stock20}.

\subsection{Priors}
\label{sec:methods_priors}

RV measurements generally do not constrain the system inclination $i$; we must therefore use an isotropic prior of $\pi(i)=\sin i$.
However, RV measurements do provide stricter priors on other parameters, particularly the minimum masses, orbital periods, and sometimes eccentricities.
Best-fitting values and uncertainties are usually available at NEA.
We therefore use the posteriors from these RV studies as priors for our stability analyses.
For the minimum masses, orbital periods, and eccentricities, we take the priors to be normal distributions with standard deviations equal to the average of the positive and negative uncertainties (and clipped where appropriate, e.g.\ at 0).
Longitudes of periastron $\Delta\varpi$ and true anomalies $\Delta f_0$ are rarely well-constrained by RV data; we therefore choose uniform priors $\pi(\Delta\varpi) = \pi(\Delta f_0) = 1/2\pi$ for these parameters.

We furthermore assume each parameter to be independent in our priors, which are thus given by
\begin{multline}
\label{eq:prior}
\pi(i, \vb\Theta)
= \pi(i)\ \pi(M_*)\ \\
    \times\prod_p \pi(M_{{\rm min}, p})\ \pi(P_p)\ \pi(e_p)\ \pi(\Delta\varpi_p)\ \pi(\Delta f_{0,p}),
\end{multline}
where the product is over the planets in the system, and trivially $\pi(\Delta\varpi_1)=\delta(\Delta\varpi_1)$ and $\pi(\Delta f_{0,1}) = \delta(\Delta f_{0,1}$) for the first planet.

Of course, a more holistic procedure would apply the stability constraint directly to the RV data, or perhaps to the RV semi-amplitude $K$.
We discuss these alternative approaches in Section~\ref{sec:discussion_alternatives}.

\subsection{Stability likelihoods -- \spock}
\label{sec:methods_likelihoods}

We evaluate stability likelihoods $\mathcal{L}({\rm stable}\  |\ i, \vb{\Theta})$ using \spock's machine-learning-based \texttt{FeatureClassifier}.
For a given system configuration (stellar and planetary masses and orbital parameters), \spock{} estimates the probability that the system is dynamically stable for $10^9$ orbits of the innermost planet \citep{tamayo20}.

The time-scale of $10^9$ orbits is valid for the planetary systems investigated in this study.
The innermost planets of all but one of these systems have periods $<8~{\rm d}$ (Table~\ref{tab:systems}); $10^9$ thus orbits corresponds to no more than $20~{\rm Myr}$.
Because these stars are not exceedingly young, we assume that the \spock{} stability likelihoods are valid.
(Indeed, these likelihoods are effectively upper bounds, making the constraints found in this work conservative; see Section~\ref{sec:discussion_inherited}.)
The exception is HD~184010, whose innermost planet HD~184010~b has a period of $287~{\rm d}$ \citep{teng22}; $10^9$ orbits thus corresponds to a longer time-scale of $0.8~{\rm Gyr}$.
However, HD~184010 is an evolved K0~III star \citep{takeda08}, with an age of $2.76^{+2.24}_{-0.95}~{\rm Gyr}$ estimated from theoretical isochrones \citep{teng22}.
We therefore conclude that all of these systems are sufficiently old that instabilities identifiable with the \spock{} classifier would have already occurred.

\subsection{Bayesian inference -- \cinemas}
\label{sec:methods_cinemas}

We have developed an open-source \textsc{pip}-installable Python package, \cinemas, which wraps the accelerated likelihood estimations from \spock{} in an MCMC inference framework, to calculate the posterior inclinations and masses of exoplanets in compact multi-planet systems, with otherwise unconstrained inclinations.
The Bayesian inference is conducted using the \textsc{emcee} package \citep{emcee}.

For a flat system of $N_p$ planets, the parameter space has a dimensionality of $5N_p$: stellar mass $M_*$; inclination\footnote{
    Note that in practice we re-parametrise in terms of $\cos i$, which is uniformly-distributed on the range $[0,1]$. This is to ensure adequate sampling at low values of $i$ (which is our primary region of interest in the parameter space).
} $i$; the minimum masses $\Mmin$, periods $P$ and eccentricities $e$ of each planet; $\Delta\varpi :=\varpi - \varpi_1$ and $\Delta f_0 := f_0 - f_{0,1}$ for each planet except the inner planet\footnote{Without loss of generality}.
Further details on the MCMC setup are provided in Appendix~\ref{app:mcmcsetup}.

\subsection{Possible undetected planets}
\label{sec:methods_undetected}

This framework can also be used to constrain the properties of undetected exoplanets.
One important prior implicit in the above methods is that the \textit{number} of planets is known with certainty.
However, the properties of undetected planets can be constrained using the \cinemas{} framework, simply by including an additional planet into the priors and likelihood functions.

We exhibit this use case with the HD 215152 planetary system.
The four planets in this system -- b, c, d, e -- have orbital periods of $5.8$, $7.3$, $10.9$, and $25.2~\rm{d}$ respectively.
There is a notable gap in between planets d and e.
We therefore apply the \cinemas{} framework to a system involving the known HD~215152 planets, as well as a hypothetical fifth planet in between d and e, labelled HD~215152~\~f.
We assume priors as follows.
Minimum mass is uniformly distributed between $0.05~M_{\earth}$ (approximately Mercury) and $5.0~M_{\earth}$.
We note that a planet with a minimum mass greater than $\approx 2.5~M_{\earth}$ would have likely had a larger RV signal size than planet c, and hence have already been detected -- we include higher minimum masses in our prior here to judge whether stronger limits on undetected planets are placed by dynamical stability or by detection thresholds.
The prior on the period of HD~215152~\~f is uniformly distributed between 12 and 24 days; the eccentricity prior is uniformly distributed between 0 and $0.3$.
The results of this extra analysis is presented in Section~\ref{sec:results_undetected}.

\section{Results}
\label{sec:results}

\subsection{Example: the YZ Ceti system}
\label{sec:results_yzceti}

\begin{figure*}
\includegraphics[width=0.9\textwidth]{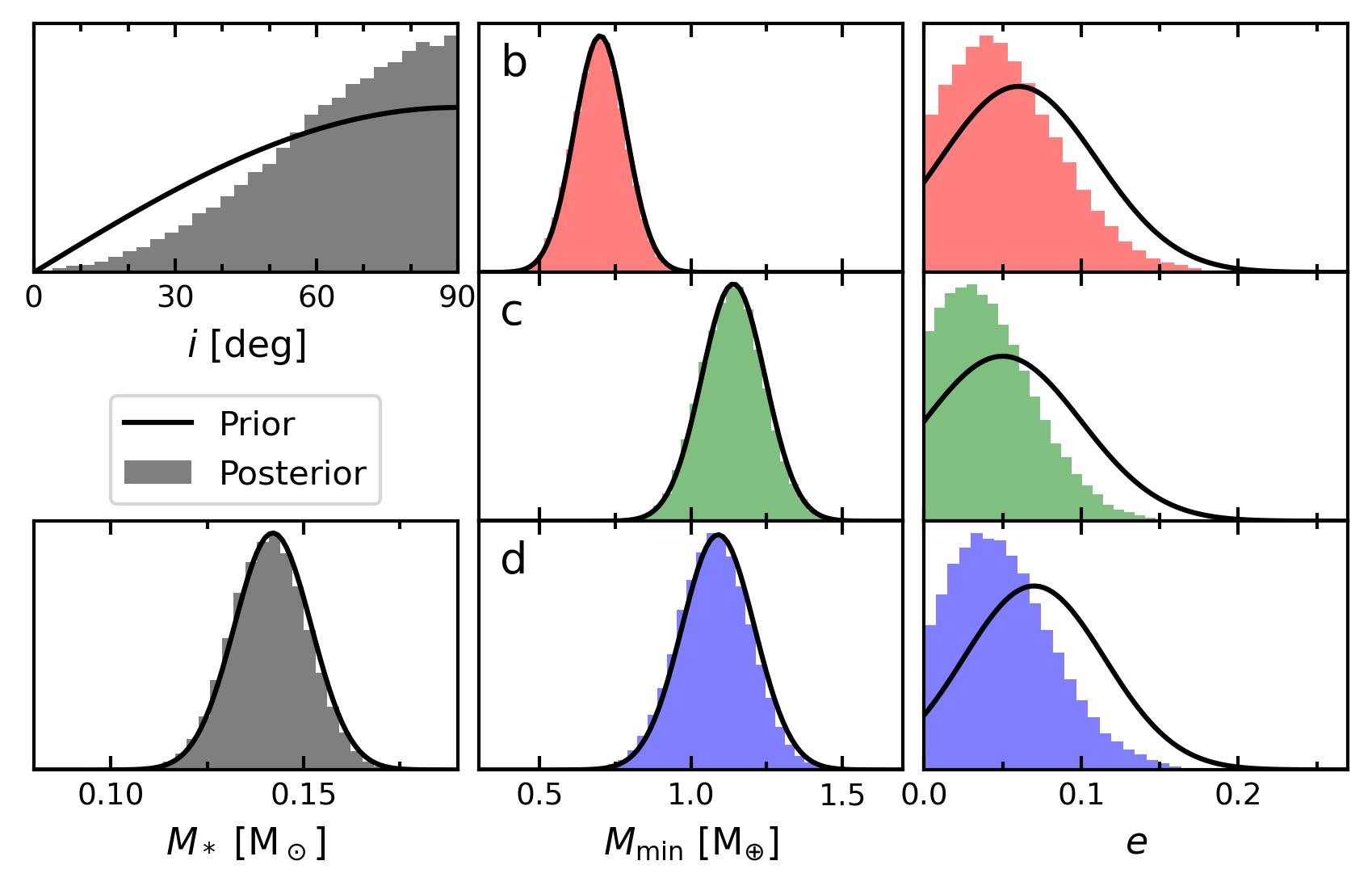}
\centering
\caption{
    Comparison of prior and posterior probability distributions for selected parameters of the YZ Ceti system.
    Lower inclinations, which imply more massive planets, are disfavoured as this makes it less likely that the system is dynamically stable (upper left panel).
    Lower eccentricities are favoured by the stability constraint (right column).
    Other parameters' posteriors, including the stellar mass (lower left) and planetary minimum masses (middle column) are not much different from the priors, suggesting that the stability constraint does not effectively constrain these parameters.
}
\label{fig:1dposteriors}
\end{figure*}

We present the three-planet YZ Ceti planetary system (Section~\ref{sec:methods_systems_yzceti}; \citealt{stock20}) as a representative example.
The MCMC walkers converge satisfactorily to the posterior distribution, which shows little covariance between parameters (see Appendix~\ref{app:fullresults} for full details).

\begin{figure}
\centering
\includegraphics[width=\columnwidth]{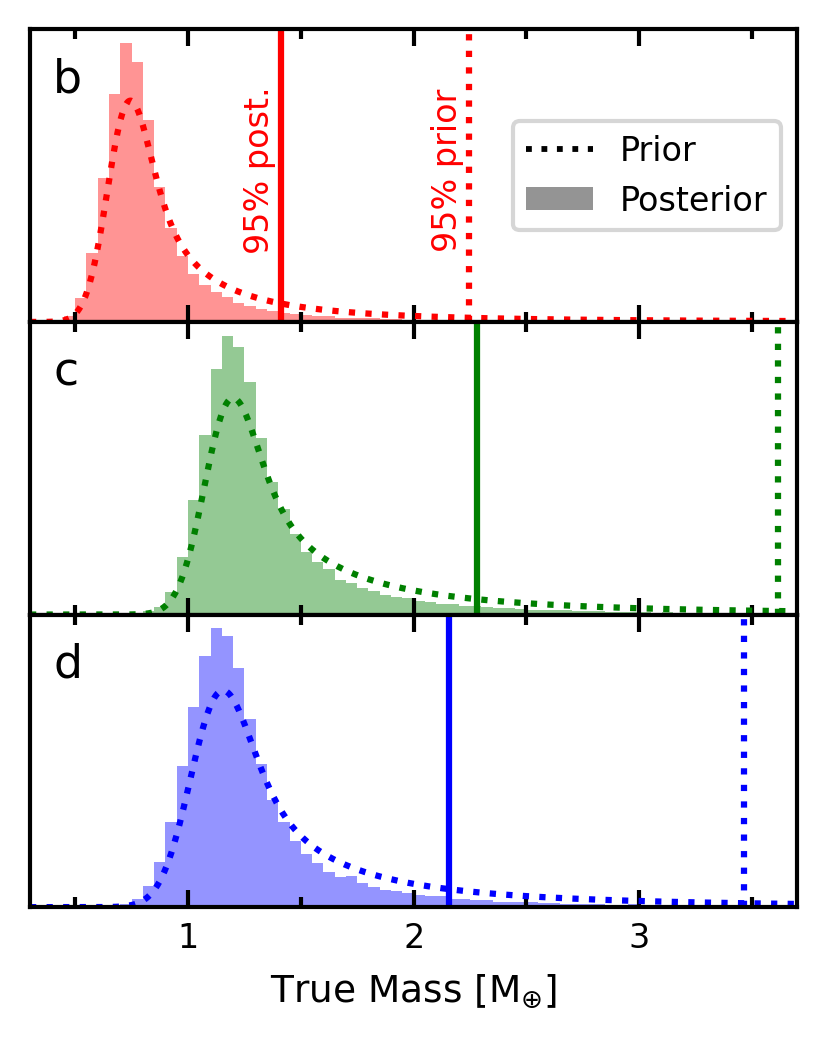}
\caption{
    Prior and posterior distributions of the true masses $M=\Mmin/\sin i$ of the YZ Ceti planets.
    The 95\% upper limits on both these distributions are shown by vertical lines.
    In each case, the two distributions look qualitatively similar, but the posterior distributions have much lower 95\% upper limits than the priors.
}
\label{fig:truemasses}
\end{figure}

The posterior distribution disfavours lower inclinations relative to the prior (Fig.~\ref{fig:1dposteriors}, upper left panel), owing to the large true masses that low inclinations would imply for its planets, and the dynamical instability this would imply for the system.
The factor by which a planet is larger than its minimum mass is $1/\sin i$; the expectation value of this factor under the prior is $\pi/2\approx 1.6$; under the posterior it is less than $1.3$.
The 95\% upper limits on the true masses are all reduced by more than one-third, and the posterior distributions are less skewed than the priors, as shown in Fig.~\ref{fig:truemasses}.
The planets are therefore more likely to have lower masses than an unconstrained inclination would suggest.

Eccentricities are also constrained, towards lower values (Fig.~\ref{fig:1dposteriors}; right column).
The priors used here do not exactly correspond to existing eccentricity uncertainties (see Section~\ref{sec:methods_priors}):
our posteriors do not necessarily represent an improvement in our knowledge of these parameters (unlike inclination, which is almost totally unconstrained \textit{a priori}).
Indeed, our eccentricity posteriors qualitatively match quite well with the $N$-body stability analyses of this system conducted in \citet[][their section~4.2]{astudillodefru17} and \citet[][their fig.~11]{stock20}.
The fact that our posterior eccentricity distributions are clearly different to our priors shows none the less that stability is an informative condition on the orbital eccentricities of planets in compact systems.

Other parameters are not effectively constrained by imposing stability.
Although there are small discrepancies, the posterior distributions on the stellar mass, planetary minimum masses, and planetary periods are very similar to the priors (Fig.~\ref{fig:1dposteriors}), implying that the likelihood function is quite flat over the prior domain of these parameters.
Additionally, the relative longitudes of periastron $\Delta\varpi$ and true anomalies $\Delta f_0$ are not constrained.

\subsection{Other systems}
\label{sec:results_others}

\begin{figure}
\centering
\includegraphics[width=\columnwidth]{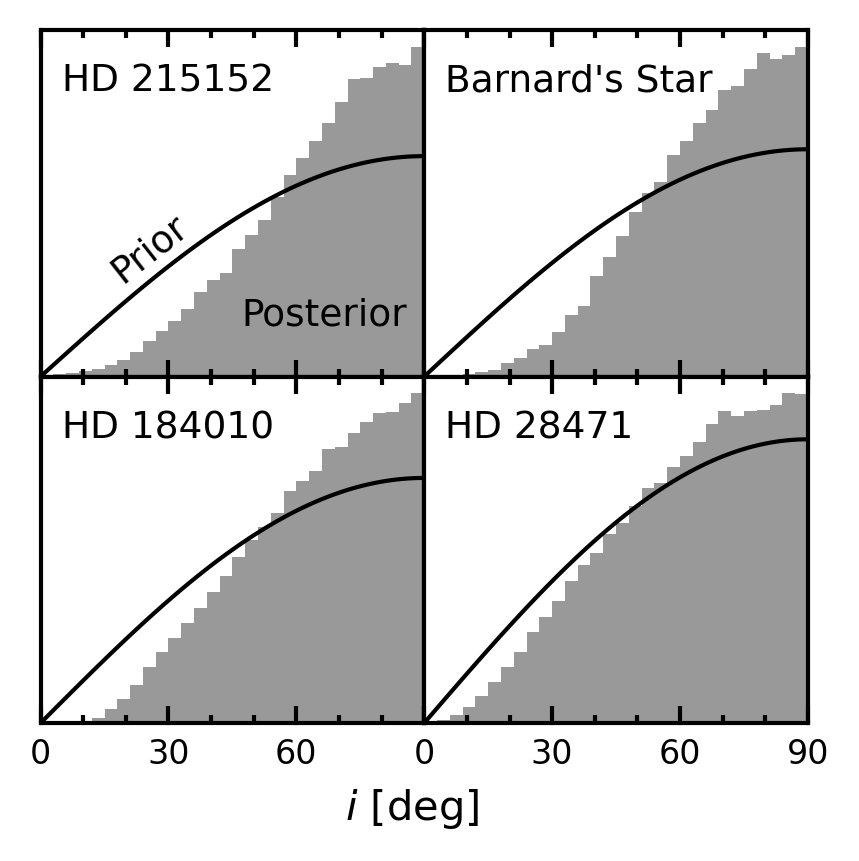}
\caption{
    Prior and posterior inclination distributions for the other four compact multi-planet RV systems.
    The distributions are presented as in Fig.~\ref{fig:1dposteriors}'s upper left panel (in which that of YZ Ceti is shown).
    The posteriors all favour (to varying degrees) higher inclinations compared to the prior, which corresponds to favouring lower masses and more dynamically stable systems.
}
\label{fig:allinclinations}
\end{figure}

Here we present the results of applying the \cinemas{} framework to all five compact multi-planet RV systems.
For each system, the posterior favours higher inclinations compared to the prior (Fig.~\ref{fig:allinclinations}), reflecting the higher stability likelihoods for the lower-mass planetary systems that these imply.
This is most starkly the case for HD~215152 and Barnard's Star, and more weakly the case for HD~184010 and HD~28471.

\begin{figure*}
\includegraphics[width=0.95\textwidth]{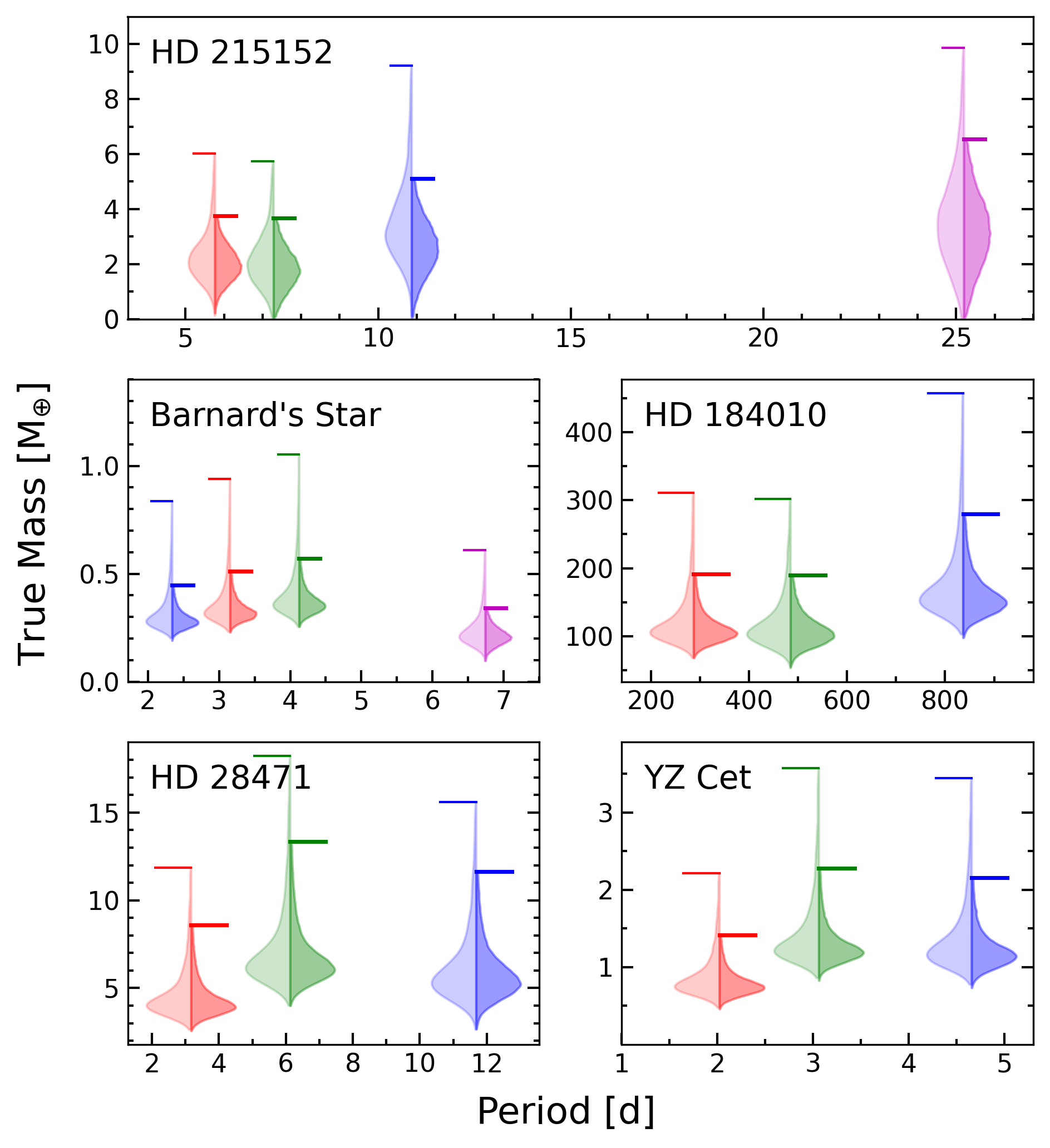}
\caption{
    Prior and posterior distributions of true mass when accounting for dynamical stability.
    The left half of each violin is the prior distribution; the right half, the posterior.
    Each distribution is truncated at the 95\% upper limit, which are labelled by horizontal lines.
    In each case the upper limits are significantly reduced when accounting for stability, due to higher masses making a system less likely to be dynamically stable.
    The planets' orbital periods are all in alphabetical order, except for Barnard's Star whose planets are named d, b, c, e \citep{gonzalezhernandez24}.
}
\label{fig:allmasses}
\end{figure*}

The prior and posterior masses for all of the planets investigated here are presented in Fig.~\ref{fig:allmasses}.
In each case, the distributions look similar overall, except for the high-mass tails, which are significantly truncated in the posterior compared to the prior.
Again, this is because higher masses are less likely to be conducive to dynamically stable systems.
To quantify the importance of accounting for stability, we present in Table~\ref{tab:masses95} the upper limits (95\%) on the planetary masses, under the prior and posterior distributions.
The upper limits on the masses are all reduced by between 20 and 45\%.

\begin{table}
\caption{
    Changes in 95\% upper limits on true planet masses in compact systems when accounting for dynamical stability.
    Planets in each system are presented in order of orbital period (this is not always alphabetical).
    The upper limits are reduced significantly; in each case between 20 and 45\%.
}
\centering
\begin{tabular}{rrrrr}
\hline
Planet & $M_{\rm min}$ & \multicolumn{2}{c}{95\% mass limit [$M_{\earth}$]} & Change \\
 & [$M_{\earth}$] & Prior & Posterior & [\%] \\
\hline
HD 215152 b & 1.82 & 5.85 & 3.75 & $-$35.8 \\
c & 1.72 & 5.69 & 3.67 & $-$35.5 \\
d & 2.80 & 9.00 & 5.11 & $-$43.3 \\
e & 2.88 & 9.72 & 6.55 & $-$32.7 \\
\hline
Barnard's Star d & 0.26 & 0.81 & 0.45 & $-$44.6 \\
b & 0.30 & 0.92 & 0.51 & $-$44.5 \\
c & 0.34 & 1.03 & 0.57 & $-$44.6 \\
e & 0.19 & 0.60 & 0.34 & $-$42.9 \\
\hline
HD 184010 b & 98.53 & 303.87 & 191.77 & $-$36.9 \\
c & 95.35 & 295.49 & 189.58 & $-$35.8 \\
d & 143.02 & 440.96 & 279.62 & $-$36.6 \\
\hline
HD 28471 b & 3.72 & 11.47 & 8.59 & $-$25.1 \\
c & 5.72 & 17.64 & 13.35 & $-$24.3 \\
d & 4.91 & 15.25 & 11.65 & $-$23.6 \\
\hline
YZ Ceti b & 0.70 & 2.16 & 1.41 & $-$34.6 \\
c & 1.14 & 3.51 & 2.28 & $-$35.0 \\
d & 1.09 & 3.36 & 2.16 & $-$35.8 \\
\hline
\end{tabular}

\label{tab:masses95}
\end{table}

\subsection{Constraints on undetected planets}
\label{sec:results_undetected}

This section presents constraints on the properties of a hypothetical planet, HD~215152~\~f, in between the planets d and e of that system (see Section~\ref{sec:methods_undetected}).

\begin{figure*}
\includegraphics[width=\textwidth]{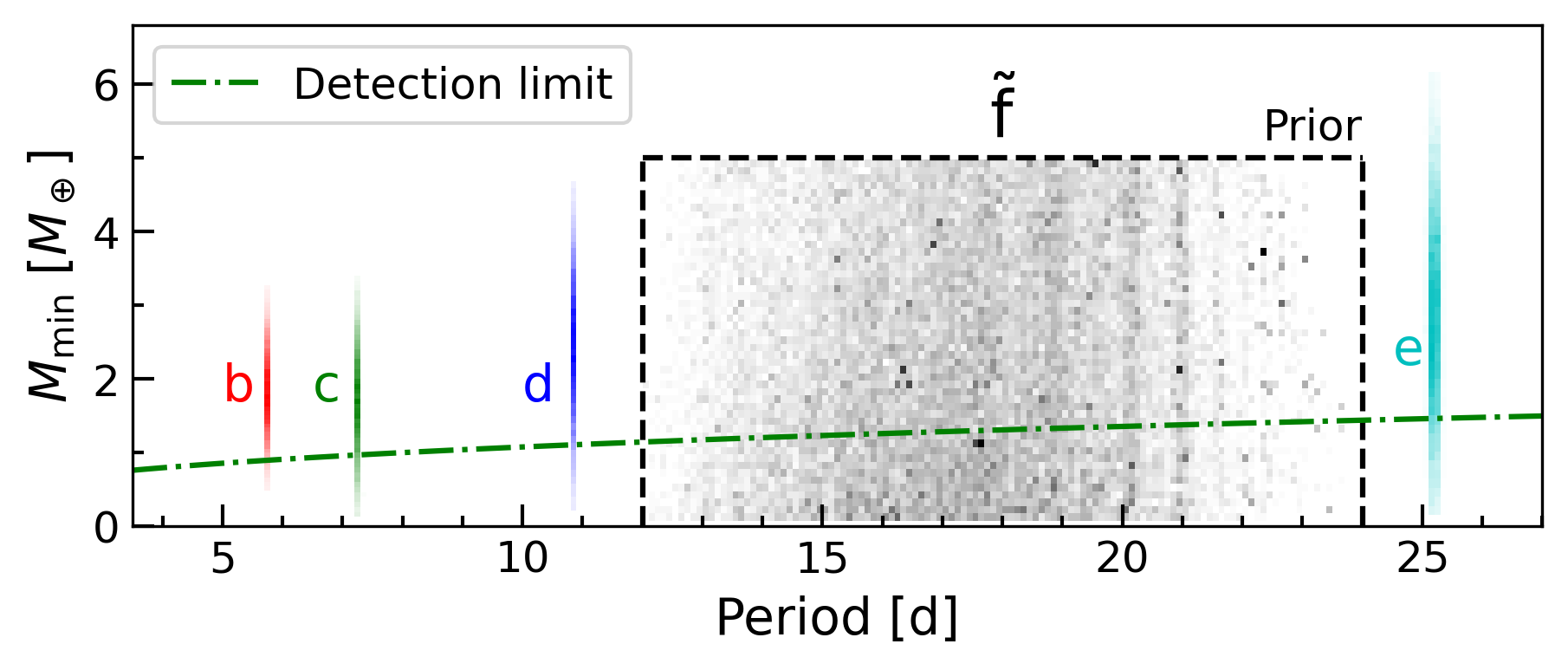}
\caption{
    Posterior distribution of $\Mmin$ and period of hypothetical planet HD~215152~\~f, calculated with the \cinemas{} framework.
    The approximate detection limit (dot-dashed line) is estimated based on planet c (see text).
    Periods too close to planets d and e are disfavoured.
    Lower minimum masses for planet \~f are slightly favoured, but values up to the top of the prior are none the less permitted.
    % The period distribution also shows faint `ripples', which may correspond to orbital resonances.
}
\label{fig:hd215152f}
\end{figure*}

The posteriors calculated with the \cinemas{} framework in $\Mmin$--$P$ space are presented in Fig.~\ref{fig:hd215152f}, along with an approximate detection limit.
Our approximation of the detection limit assumes that:
(a) RV semi-amplitude $K$ alone determines detectability;
(b) $K \propto M_{\rm min} P^{-1/3}$; and
(c) the lowest detectable signal corresponds to planet c, at its $-1\sigma$ minimum mass.

The posteriors are surprisingly permissive to large minimum masses.
Although smaller minimum masses are slightly favoured, large values are permitted up to the top of the prior ($\Mmin=5~M_{\earth}$).
Such planets would certainly have been seen in RV observations, as their radial velocity amplitude would be significantly greater than that of planet c.
As such, for the detection of this hypothetical planet, the restrictions imposed by injection-recovery tests would be more informative than stability analyses.

Our framework is more valuable in restricting the possible \textit{periods} of such a hypothetical planet.
Periods too close to planets d and e are disfavoured in the posterior distribution (Fig.~\ref{fig:hd215152f}).
Inevitably such nearby orbits are more likely to give rise to instabilities.
% A more subtle feature of the period distribution is the `rippling' seen between 20 and 22 days.

\section{Discussion}
\label{sec:discussion}

\subsection{Implications}
\label{sec:discussion_implications}

% The results above limit the masses of exoplanets in certain systems.
% \textit{A priori}, RV exoplanets could have almost arbitrarily high masses, to the point that they could in fact not be exoplanets at all; but rather brown dwarfs or even low-mass stars \citep{kiefer21}.
% However, in compact multi-planet systems such as those studied here, such high masses would not be dynamically stable.
% For instance, if YZ Ceti's planets were all in fact brown dwarfs, the compact system would have undergone dynamical instabilities long ago.

Our results limit the masses of exoplanets in certain systems, which is vital for interpreting the nature of these planets.

The YZ~Ceti planets have minimum masses between $0.7$ and $1.14~M_{\earth}$, but because their inclinations are unknown, their true masses could \textit{a priori} be in the sub-Neptune, or even gas giant regime.
While the distinction between super-Earths and sub-Neptunes is primarily based on radii (which are unconstrained here) and the distinction in terms of masses is blurry and somewhat stellar-mass-dependent \citep[e.g.][]{parc24}, our work strongly disfavours sub-Neptune interpretations for the planets in the YZ Ceti and Barnard's Star systems.

We furthermore strengthen the likelihood that the planets of HD~2151512 are super-Earths rather than sub-Neptunes: while their prior true masses are below $10~M_{\earth}$, we constrain them further to be below $7~M_{\earth}$ (Table~\ref{tab:masses95}).
While the high-mass tail here allows for sub-Neptunes, lower inclinations are less probable \textit{a priori} and even more so \textit{a posteriori} (Fig.~\ref{fig:allinclinations}).
Our work therefore makes sub-Neptune interpretations of these planets even less likely, especially for the smaller planets b and c.

The nature of the planets in the HD~28471 system is less well-constrained by our framework, though this is ultimately due to the overlap in mass distributions of super-Earths and sub-Neptunes.
The minimum masses of the planets in this system are between 3 and $6~M_{\earth}$ (Table~\ref{tab:systems}).
While we rule out masses greater than $14~M_{\earth}$ for all these planets, and while lower masses are even more favoured than \textit{a priori}, the posterior distributions of these planets' masses do allow for both super-Earth and sub-Neptune interpretations.

Finally, the gas giants in the HD~184010 are all constrained to be intermediate in mass between Saturn and Jupiter.
Their minimum masses are between $1.00$ and $1.50$ Saturn masses, and the prior mass distributions allowed for planet d and potentially planet c to be of Jupiter mass or greater.
The dynamical constraints in our work limit the mass of even the most massive planet (d) to be at most $0.88~M_J$ (Table~\ref{tab:masses95}).

\subsection{Comparisons with numerical simulations}
\label{sec:discussion_numerical}

\subsubsection{Mutual Hill radii}
\label{sec:discussion_numerical_hill}

Here we compare our findings with the results of a suite of numerical simulations from the seminal work of \citet{chambers96}.
Therein it is found that systems in which planets whose semi-major axes are less than 10 mutual Hill radii apart -- where the mutual Hill radius for planets labelled 1 and 2 is given by
\begin{equation}
R_{\rm H}
= \qty(\frac{M_1 + M_2}{3M_*})^{1/3} \frac{a_1 + a_2}{2}
\end{equation}
-- are always unstable.
This criterion can expressed equivalently as $\Delta>10$, where
\begin{align}
\Delta
\equiv \frac{a_2-a_1}{R_{\rm H}}
&= 2 \qty(\frac{a_2-a_1}{a_1 + a_2})
    \qty(\frac{3M_*}{M_1 + M_2})^{1/3}\\
\label{eq:hill_inclination}
&= 2 \qty(\frac{a_2-a_1}{a_1 + a_2})
    \qty(\frac{3M_*}{M_{{\rm min,}1} + M_{{\rm min,}2}})^{1/3} \qty(\sin i)^{1/3}
\end{align}
This criterion was uncovered under the assumptions of a solar-mass star, specific (and equal) planetary masses, or pseudo-uniformly spaced semi-major axes.
The systems under investigation here obviously do not align with these assumptions, but this `$\Delta>10$' criterion provides an interesting fiducial comparison none the less.

\begin{figure}
\includegraphics[width=\columnwidth]{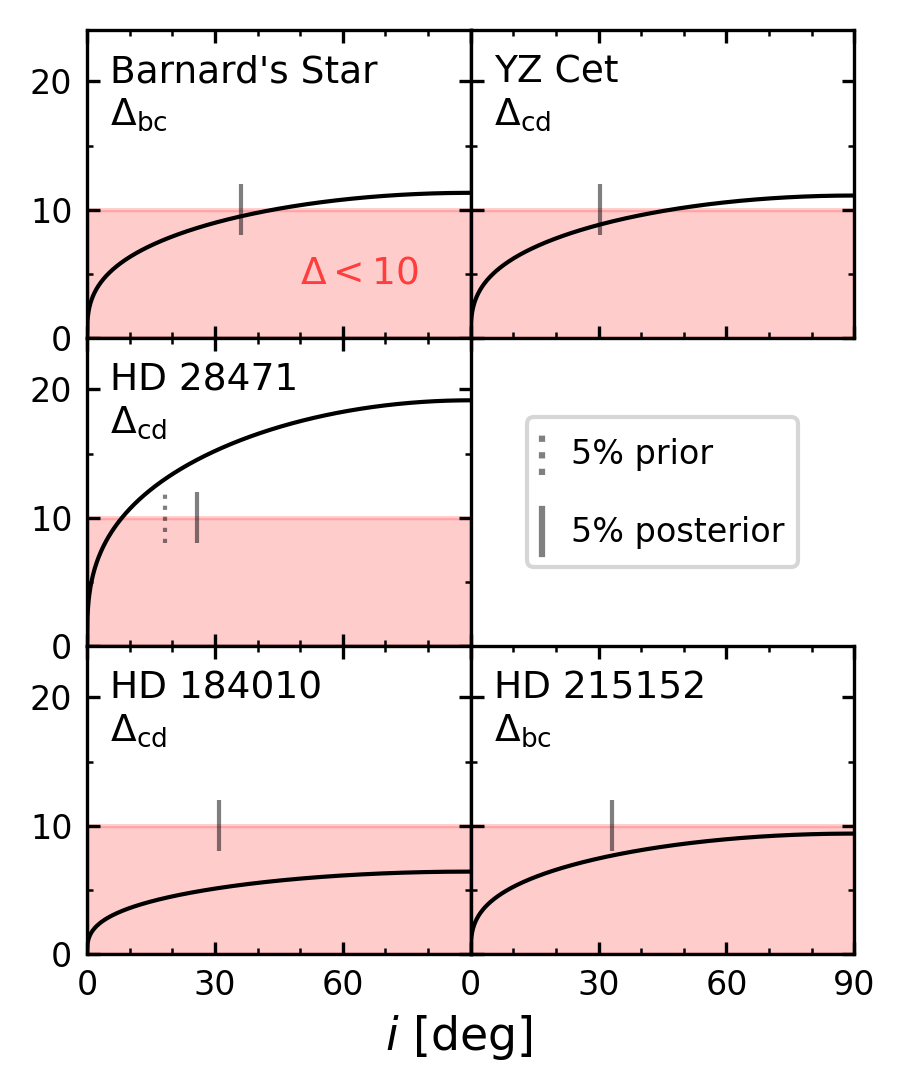}
\caption{
    Minimum separation of planets' semi-major axes in mutual Hill radii ($\Delta$), as a function of inclination ($i$).
    The physical separation between planets' semi-major axes is known, but the mutual Hill radius depends on masses, and hence on $i$ ($\Delta\propto(\sin i)^{1/3}$; equation~\ref{eq:hill_inclination}).
    Each curve corresponds to the pair of planets with the smallest Hill separation.
    The red region ($\Delta<10$) is unstable according to the criterion from \citet{chambers96}.
    The solid vertical lines show the 5\% lower inclination limits calculated using \cinemas{} for each system, corresponding to the 95\% upper mass limits.
    For Barnard's Star and YZ Cet, the \cinemas{} limits are close to the inclination at which the curve crosses the $\Delta>10$ threshold.
    For HD~184010 and HD~215152, the planets are separated by $\Delta<10$ even at $i=90\degr$ (corresponding to minimum masses).
    Conversely, the HD~28471 planets are separated by $\Delta>10$ even for inclinations much lower than the limits placed by \cinemas, due to the prior (dashed line; see text).
}
\label{fig:hillstability}
\end{figure}

We would expect our 95\% upper limits to be close to the $\Delta>10$ criterion: this would signify agreement between this criterion and calculations from \spock, that masses exceeding this criterion would lead to dynamical instability.
The 5\% lower limits on the inclination (equivalent to the 95\% upper limits on the masses) deduced in Section~\ref{sec:results_others} are compared to this criterion in Fig.~\ref{fig:hillstability}, and discussed here.
The comparison is different for different systems.

For the planets in the Barnard's Star and YZ Ceti systems, the \cinemas{} results agree quite well with the $\Delta>10$ criterion.
For these cases, the planet pair separated most closely in Hill radii falls below $\Delta=10$ at around the same inclination angle as our 5\% lower limit.
If the inclination were lower, the planets would be more massive, and the planets would be separated by less than $10~R_{\rm H}$.

The HD~184010 and HD~215152 systems show that the $\Delta>10$ criterion is not universally followed.
In these systems, at least one pair of planets is separated by $\Delta<10$, even if $i=90\degr$ (that is, if the masses equal the minimum masses).
These systems emphasise the heuristic nature of general criteria such as that of \citet{chambers96}, which are derived under particular assumptions.

Conversely, the planets in the HD~28471 system have $\Delta>10$ even for inclination angles much \textit{lower} than the 5\% limits calculated in \cinemas{}.
This is essentially a prior effect: the 5\% lower limit on the isotropically-distributed inclination is $\arccos(0.95)=18.2\degr$, not much smaller than the posterior value of $25.7\degr$.
Because the constraints for this system are particularly weak (Fig.~\ref{fig:allinclinations}), the 5\% posterior limit is not much higher than the prior limit.
As such, even though the system could be dynamically stable at lower inclinations (i.e.\ higher masses), \cinemas{} accounts for the fact that low inclinations are improbable \textit{a priori}, deeming them unlikely even though they would still correspond to stable systems.
Indeed, an RV system with planets on widely-spaced orbits would similarly be dynamically stable for inclinations significantly lower than limits calculated using \cinemas{}; such inclinations are simply geometrically improbable.
The estimation of the inclinations of RV systems should be understood under the important prior context that lower inclinations are less probable geometrically.

\subsubsection{Simulations of specific systems}
\label{sec:discussion_numerical_specific}

While the MCMC approach we have used here provides posterior samples, it does not provide the Bayesian \textit{evidence}.
As such, \cinemas{} does not on its own provide reassurance that a given system is stable under \textit{any} prior configuration.
One could imagine that \cinemas{} is simply returning low stability probabilities for \textit{all} configurations, with some merely being lower than others.
This would indicate (assuming the stability classification is robust) that the results of the RV observations on which our work is based had not accounted for dynamical stability in their determination of the orbital parameters of the system.

However, each of these systems \textit{has} undergone preliminary dynamical analysis, which in each case confirms that the configurations they find are not obviously unstable.
\citet{delisle18} perform $N$-body integrations of the HD~215152 system for $1~{\rm kyr}$, finding at least some stable regions of configuration space.
\citet{teng22} similarly integrate realisations of the HD~184010 system for $10~{\rm Myr}$, finding a region of eccentricity-inclination space for which the planets are all stable over that time-scale, despite their only being separated by $\Delta \approx 6.5$ even at minimum mass.
Although this is much lower than the limit from \citet{chambers96}, it is found in \citet{marzari02} that systems of three \textit{giant} planets ($M=10^{-3}M_*$) can be long-term ($\sim$Gyr) stable even for $\Delta=5.3$; the integrations in \citet{chambers96} only go up to $M=10^{-5}M_*$.
Regarding Barnard's Star, \citet{basant25} investigate the stability of this system with \spock, but \citet{byrne26} also perform a single $N$-body simulation assuming the planets had masses $2.5\times$ their minima, finding it highly stable.
As part of the determination of the orbital parameters of the HD~28471 planets in \citet{stevenson25}, the angular momentum deficit (AMD; \citealt{laskar97, laskar00, laskar17, petit17}) is employed, outright rejecting configurations for which this deficit implies instability (as part of the \textsc{kima} framework; \citealt{faria18}).
\citet{stock20} employ a similar AMD-based approach to downweight unstable configurations of the YZ Ceti system.
In both cases, stable configurations remain.

These existing analyses of the systems' dynamical stabilities are not comprehensive, but they reassure us that each system has at least some stable region of parameter space, justifying our MCMC procedure.

\subsection{Alternative approaches}
\label{sec:discussion_alternatives}
Our approach towards constraining the true masses of exoplanets using dynamical stability is not the most holistic possible approach.
Here we discuss certain alternative possible frameworks.

\subsubsection{RV semi-amplitude}
\label{sec:discussion_alternatives_K}

Technically, RV measurements do not directly measure the minimum mass, but rather the RV semi-amplitude:
\begin{equation}
\label{eq:K}
K = \qty(\frac{2\pi G}{P})^{1/3} \qty(M_*+M)^{-2/3} \qty(1-e^2)^{-1/2} \times \qty(M \sin i),
\end{equation}
where we see that the true mass $M$ also enters into the semi-amplitude via the $(M_*+M)^{-2/3}$ term, which is only independent of $M$ in the limit $M\ll M_*$.

Given that our results disfavour large masses, it is necessary to check the validity of this approximation.
Assuming $M\ll M_*$ effectively imparts a biasing factor of $\qty(1 + M / M_*)^{-2/3}$ towards lower masses.
For our posterior distributions, we find that this factor exceeds unity by no more than $10^{-3}$ for HD~184010, and no more than $10^{-4}$ for all the other systems.
This test is not fully self-consistent, but it gives us confidence that the approximation of $M\ll M_*$ is valid for stable configurations of the systems investigated here.
None the less, this may not be the case for all future systems to which this framework could be applied, and an alternative framework which directly incorporates $K$ rather than $\Mmin$ would be more holistic.

\subsubsection{RV measurements}
\label{sec:discussion_alternatives_RVs}

Yet more holistic would be to use the radial velocity measurements directly, rather than indirectly via the distributions of $\Mmin$, $P$ etc.\ that are derived from them.
Such a framework would be similar to the \textsc{kima} framework, in which AMD-stability can be enforced in the inference of orbital parameters from RV data \citep{faria18}.
However, the use of \textsc{spock} stability likelihoods, rather than binary AMD stability, would have the advantage of more smoothly incorporating the probabilistic nature of dynamical stability.

We do not present a fully holistic framework here, in our attempt to present the concept of Bayesian stability-based inclination constraints as plainly as possible.
Incorporating the principles of \cinemas{} into an RV analysis pipeline would -- although beyond the scope of this work -- doubtless be a valuable addition the pipeline.

\subsection{Future work}
\label{sec:discussion_future}

The flexibility of the \cinemas{} framework suggests further applications, two of which are outlined here.

\subsubsection{Model comparison}
\label{sec:discussion_future_modelcomparison}

The MCMC implementation outlined above is not capable of preferring configurations with different numbers of planets, even though there is inevitably a stability differential between such configurations.
This is because MCMC approaches do not generally calculate the Bayesian evidence for a given configuration; only samples from the posterior distribution.
This contrasts with numerous exoplanet discovery studies, in which the evidence for models with different numbers of planets is often compared.

These model comparisons typically do not incorporate the stability differential; systems with more planets are generally less stable than those with fewer planets.
Incorporating stability probabilities into these model comparisons would alter the degree of preference for one model over another (most likely towards fewer-planet solutions).

Such an incorporation of stability likelihoods into model comparison could be implemented with the \cinemas{} framework if, rather than using an MCMC approach, a nested sampling approach is used instead \citep{skilling04}.
Nested sampling \textit{does} provide an evidence, facilitating quantitative comparisons between models with different numbers of planets, in a manner that does not ignore dynamical (in)stability.
Nested sampling is furthermore implemented as an alternate inference mode in the \cinemas{} package, though we have only used the MCMC mode here.
Alternatively, the Learned Harmonic Mean Estimator (LHME; \citealt{mcewen23, polanska25}) could be used to calculate the evidence directly from the MCMC samples \citep{dobson26}

\subsubsection{Outer giant planets}
\label{sec:discussion_future_outergiants}

Aside from HD~184010, all the planetary systems investigated here contain a number of small, close-in planets.
On a population level, such systems are often accompanied by outer giant planets (\citealt{zhu18, bryan19, rosenthal22}; but see also \citealt{bonomo23}).
Several thousand planets of a few Jupiter masses at $\sim1~{\rm au}$ are expected to be detected in the \textit{Gaia} mission's fourth data release (DR4; \citealt{perryman14, wallace25, lammers26}), due to be released in December 2026.

A clear additional application of the \cinemas{} framework is to determine stable orbital parameters of giant planets orbiting further out.
This could be done analogously to our constraining of the parameters of an additional planet within the HD~215152 system (Section~\ref{sec:methods_undetected}).
We find, however, that for the systems studied here, a Jupiter-mass planet on periods $\sim 10^4~{\rm d}$ would have an RV signal size significantly larger than that of planets already detected, and can therefore be ruled out anyway.
Thus, in these cases, the properties of such planets would be constrained much more effectively by injection-recovery tests than by dynamical stability considerations, though we note that this may not be the case for systems identified in the future.
This mirrors our findings from our search for HD~215152~\~f within the known inner system (Section~\ref{sec:results_undetected}), in which it is found that injection-recovery tests would provide much stricter mass constraints than stability considerations.

\subsection{Limitations}
\label{sec:discussion_limitations}

\subsubsection{Scope}
\label{sec:discussion_limitations_scope}

The scope of these results is limited to particular systems.
Firstly, we do not consider transiting systems, which have $i\approx 90\degr$ and therefore the true mass is effectively equal to the minimum mass.
Secondly, we do not consider single-planet systems, which of course have no planet-planet interactions, and are dynamically stable for arbitrarily large masses.
The stability of two-planet systems is furthermore analytically solvable \citep{wisdom80, deck13, hadden18}.

Finally, in multi-planet systems, a certain level of compactness is required for planet-planet interactions to be important, thus enabling dynamical constraints on the masses
\citep[e.g.][]{wisdom80, chambers96, quillen11}.
In systems which are less dynamically compact than those studied here, planet-planet interactions will be weaker, and hence only very low inclinations could be ruled out.
One way of quantifying the compactness of a system is using the characteristic spacing $\mathcal{S}$ \citep{gilbert20}, equal to the mean separation of the planets in units of mutual Hill radii where $M_*$ is the stellar mass, $M_1$ and $M_2$ are the planets' masses, and $a_1$ and $a_2$ their semi-major axes.
By this metric, the least compact system in our sample is HD~28471, with $\mathcal{S}=19.7$; indeed the masses of the planets in this system are constrained more weakly than for other systems (Table~\ref{tab:masses95}).
We would expect that \cinemas{} would give even weaker mass constraints on more sparsely-spaced systems than those investigated here.

\subsubsection{Mutual inclinations}
\label{sec:discussion_limitations_mutualinclinations}

Mutual inclinations between the orbits of planets in the same system are not considered in this work.
Our reasons for doing so are primarily practical, but there are observational justifications for this omission.

Going from a flat system to a mutually-inclined system requires two extra parameters per planet\footnote{
    Technically, these extra parameters would not be needed for the innermost planet without loss of generality.
}: the mutual inclination itself ($\Delta i$), and the longitude of the ascending node ($\Omega$).
This would take the dimensionality of the parameter space from $5N_p$ to $\approx7N_p$.
MCMC -- as well as other Bayesian inference strategies such as nested sampling \cite{skilling04} -- suffer from the so-called `curse of dimensionality', with computation time and convergence deteriorating in parameter spaces of higher dimensionality \citep[e.g.,][]{gelman97,roberts01,skilling09,chopin10}.
This computational cost similarly makes quantitatively testing the validity of the flat-system assumption prohibitive for the purposes of this proof-of-concept work.

However, if the planets have \textit{small} mutual inclinations, this would not significantly affect the implied masses.
If a system has an average inclination of $60\degr$, but one of its planets in fact has an individual inclination of $55\degr$ (corresponding to a relatively large mutual inclination of $5\degr$), this would only alter the inferred mass by 5\%.
Given that our posterior distributions are not tight anyway (our focus is on tightening the upper mass limits), we do not see the additional accuracy resulting from incorporating mutual inclinations as worth the extra computational expense.

Fortunately, systems with large mutual inclinations are observationally rare.
Transit timing and duration variations (TTVs and TDVs) show that the multi-planet systems observed by the \textit{Kepler} mission \citep{kepler} are generally flat within a few degrees \citep[e.g.][]{lissauer11,fang12,fabrycky14}.
Consistency arguments between RV and transit surveys corroborate this conclusion \citep{figueira12}.
Individual systems with large mutual inclinations have been found -- from TTVs \citep[e.g.][]{nabbie25}, astrometry \citep[e.g.][]{derosa20} or dynamical arguments \citep{dawson14} -- but surveys of short-period planetary systems show such systems to be statistically exceptional.
(We note however that the distribution of mutual inclinations of longer-period planetary systems, such as HD~184010, is less well-constrained.)

The systems we investigate here (with the possible exception of HD~184010) are thus unlikely to have significant mutual inclinations; but even if they were, the implications for the masses are likely to be minor, meaning that the errors on the inferred posterior mass distributions are unlikely to be significant.
This could of course be investigated by incorporating the relevant extra parameters into the inference framework, but we do not see this as worth the required additional computational cost.
We therefore deem mutual inclinations beyond the scope of this work, but note that their incorporation could none the less influence dynamical stability.

\subsubsection{Limitations inherited from upstream software}
\label{sec:discussion_inherited}

This work is built upon the \spock{} stability classifier, which is in turn built upon the \rebound{} integrator \citep{rebound}.
Our results are therefore valid up to the validity of these upstream codes.
Here we discuss some limitations of these codes, and how they may impact our results.

Firstly, instabilities on long time-scales are not considered by \spock{}, though this likely means the constraints presented in this work are conservative.
\spock{} determines the probability of dynamical stability over $10^9$ orbits of the innermost planet \citep{tamayo20}.
However, the systems considered here are much older than this time-scale (Section~\ref{sec:methods_likelihoods}; Table~\ref{tab:systems}).
Some regions of parameter space might therefore be unstable on time-scales of, say, $1.1\times10^9$ orbits, but not $1.0\times10^9$ orbits.
Such regions would not be excludable by \spock, even if parameters in this region are impossible for systems older than this.
If we assume that dynamical stability is, vaguely speaking, a monotonically decreasing function of mass, then it is the higher-mass end of the distributions that would be excluded by instabilities on longer time-scales.
As such, if \spock{} were sensitive to longer time-scales, our posteriors would be expected to be \textit{even more} skewed towards lower masses.

Second, it is possible that the size of the training set for \spock{} could be insufficient to sufficiently sample the parameter space.
Due to the inherently chaotic nature of multi-body gravitational systems, unstable regions of parameter space may not be sufficiently well-sampled by the training set.
However, \citet{tamayo20} none the less find the \spock{} classifier to be highly robust and to generalize well, validating against $N$-body simulations with an area-under-curve (AUC) score of 0.95 \citep{tamayo20}.
Additionally, it was found in \citet{byrne26} that the \spock{} stability classification agrees well with $N$-body simulations for the case of the Barnard's Star planetary system.
We are therefore confident that \spock{} provides an accurate estimate of the stability probability for various configurations of the systems investigated here.

Finally, we note that the first step of \spock's stability classification is to numerically integrate the system for $10^4$ orbits using \rebound; if the integration were inaccurate, this would therefore compromise the accuracy of the stability calculations.
For example, general-relativistic (GR) precessional effects can significantly affect -- both positively and negatively -- the stability of compact planetary systems \citep[e.g.][]{adams06a, adams06b, naoz13, wei21}.
The importance of GR compared to secular interactions (between planets labelled 1 and 2) can be quantified by the dimensionless parameter $\Pi_0$ \citep{adams06a}, where
\begin{equation}
\label{eq:gr}
\Pi_0 = \frac{4G M_*^2 a_2^3}{M_2 c^2 a_1^4}
\end{equation}
We test the importance of GR in a similar fashion to our test of the importance of the mass biasing factor in Section~\ref{sec:discussion_alternatives_K}, by evaluating this correction term on our posterior samples.
We find that for our systems this correction term never exceeds $3\times 10^{-4}$.
Again, this is not a fully self-consistent test, as these samples were generated without incorporating GR.
None the less, we are confident that the results presented in this work would be robust to GR effects.
The application of \cinemas{} to ultra-compact systems in which GR-induced precession is more important is however inadvised.

\subsubsection{Giant planets}
\label{sec:discussion_limitations_giants}

One of the systems investigated in this work, HD~184010, contains three giant planets ($\Mmin\sim100~M_{\earth}$).
However, the simulations that \spock{} was trained on have a maximum planet-star mass ratio of $\approx 2M_{\rm Neptune}/M_{\sun} \approx 35 M_{\earth}/M_{\sun}$ \citep{tamayo20}.
Applying \cinemas{} to the HD~184010 system therefore constitutes extrapolation beyond the distribution of the training data used for \spock.
As such, the stability probabilities produced by the \spock{} classifier may be inaccurate.
Although our results for the HD~184010 system agree with the $N$-body simulations of this system in \citet{teng22}
-- we similarly find the system to be unstable for inclinations $\lesssim 15^\circ$ (Fig.~\ref{fig:allinclinations}) --
we none the less recommend caution when applying our framework to giant planets, due to limitations in the stability likelihood calculations.

% Something about resonances??

\section{Conclusions}
\label{sec:conclusion}

We have presented \cinemas{}, a Bayesian framework for constraining the true masses of RV-detected exoplanets in flat multi-planet systems, whose masses are otherwise degenerate with the inclination.
Applying this framework to five compact systems, we cut the 95\% upper mass limits by between 20 and 45\%, ruling out high masses previously permitted by unconstrained inclinations.
We also apply this framework for determining the parameters of undetected planets, though stricter constraints are usually provided by injection-recovery tests.
Although transiting exoplanetary systems are much easier to study than RV-discovered systems, they are geometrically highly improbable, meaning that methods such as ours that target non-transiting systems are crucial for the study of nearby exoplanets -- as well as the thousands of non-transiting planets expected to be discovered in \textit{Gaia} DR4.

\section*{Acknowledgements}

We thank two anonymous reviewers, whose comments greatly improved many aspects of this work.

We acknowledge productive conversations with Lorenzo Pica~Ciamarra, Dimitri Veras, and Claire McLellan-Cassivi.

In addition to Python packages referenced in the text, we also acknowledge the use of \textsc{Astropy} \citep{astropy1, astropy2, astropy3}, \textsc{corner} \citep{corner}, \textsc{Matplotlib} \citep{matplotlib}, \textsc{NumPy} \citep{numpy}, and \textsc{pandas} \citep{pandas1, pandas2}.

This research has made use of the NASA Exoplanet Archive, which is operated by the California Institute of Technology, under contract with the National Aeronautics and Space Administration under the Exoplanet Exploration Program.

This work has made use of data from the European Space Agency (ESA) mission {\it Gaia} (\url{https://www.cosmos.esa.int/gaia}), processed by the {\it Gaia} Data Processing and Analysis Consortium (DPAC, \url{https://www.cosmos.esa.int/web/gaia/dpac/consortium}).
Funding for the DPAC has been provided by national institutions, in particular the institutions participating in the {\it Gaia} Multilateral Agreement.

The authors declare no conflicts of interest.

%%%%%%%%%%%%%%%%%%%%%%%%%%%%%%%%%%%%%%%%%%%%%%%%%%
\section*{Data Availability}

The data used in this work are all publicly available, mostly from the NASA Exoplanet Archive.
The Python package we have developed, \cinemas{}, is freely-available to install using \textsc{pip}, or at \url{https://github.com/xbyrne/cinemas}.
Scripts and samples for the MCMC runs and post-analysis are available at \url{https://github.com/xbyrne/inclinations}.



\bibliographystyle{rasti}
\bibliography{bibliography}

%\begin{thebibliography}{99}
%\end{thebibliography}

%%%%%%%%%%%%%%%%%%%%%%%%%%%%%%%%%%%%%%%%%%%%%%%%%%

%%%%%%%%%%%%%%%%% APPENDICES %%%%%%%%%%%%%%%%%%%%%

\appendix

\section{Distributions of minimum and true masses}
\label{app:truevmin}

We derive here relationships between the probability distribution of the minimum mass $\Mmin$ (obtained e.g.\ from RV data) and the corresponding distribution of the true mass $M=\Mmin/\sin i$.
Note that here we are considering the probability distributions for a single planet; not the distribution of true masses among the exoplanet population (as in e.g.\ \citealt{zucker01, jorissen01}).

Assuming inclinations are distributed isotropically, we have $\Pr(i)=\sin i$.
Thus
\begin{equation}
\Pr(M)\dd{M} = \int_0^{\pi/2} \dd{i} \dd{\Mmin}(i) \Pr(\Mmin, i) .
\end{equation}
Using $\dv*{\Mmin}{M}=\sin i$, and assuming that the minimum mass is independent of the inclination such that $\Pr(\Mmin, i) = \Pr(\Mmin)\Pr(i)$, we obtain
\begin{equation}
\Pr(M) = \int_0^{\pi/2} \dd{i} \Pr(\Mmin) \sin^2 i.
\end{equation}
For example, if $\Mmin$ follows a Gaussian distribution with mean $\mu$ and variance $\sigma^2$, this distribution would be given by
\begin{equation}
\Pr(M) \propto
    \int_0^{\pi/2} \dd{i}
    \exp\qty[
        -\frac{1}{2} \qty(\frac{M \sin i - \mu}{\sigma})^2
    ]
    \sin^2 i.
\end{equation}

We now show that the expectation value of the true mass is greater than that of the minimum mass by a factor $\pi/2$, irrespective of the form of $\Pr(\Mmin)$.
The expected value of the true mass is given by
\begin{align}
\mathbb{E}[M] &\equiv \int_0^{\infty} \dd{M} M \Pr(M)\\
&= \int_0^\infty \dd{M} M \int_0^{\pi/2} \dd{i} \Pr(\Mmin=M\sin i) \sin^2 i.
\end{align}
Swapping the order of integration and using the substitution $\Mmin=M\sin i$, we obtain
\begin{equation}
\mathbb{E}[M] = \int_0^{\pi/2} \dd{i} \int_0^\infty \dd\Mmin \Mmin \Pr(\Mmin)
= \frac{\pi}{2} \times \mathbb{E}[\Mmin]
\end{equation}
For the specific case that the minimum mass is known exactly to be, say, $\Mmin=M_0$, the expected true mass is then $M\approx1.57M_0$.

The \textit{variance} of the true mass, however, is formally undefined.
If we attempt to take the expectation value of $M^2$, we arrive at
\begin{equation}
\mathbb{E}[M^2] = \int_0^{\pi/2} \frac{\dd{i}}{\sin i} \times \mathbb{E}[\Mmin^2],
\end{equation}
the integral over $i$ diverging to infinity, as therefore does the variance.

\section{MCMC setup}
\label{app:mcmcsetup}

The parameter space of a flat system of $N_p$ planets has a dimensionality of $5N_p$, as outlined in Section~\ref{sec:methods_cinemas}.
We run each MCMC for 20\,000 links, using the \textsc{emcee} package \citep{emcee}.
We use a number of walkers equal to three times the number of parameters ($15N_p$).
The proposal generators used were differential evolution \citep{demove} with a stretch factor $\gamma_0=0.2$, and differential evolution with snooker updater \citep{desnookermove}; these generators were weighted 0.9 and 0.1 respectively.

The initial states were generated as follows:
\begin{itemize}
\item $\cos i$ is generated uniformly between 0 and 1;
\item stellar mass, planetary minimum masses, and periods are sampled from normal distributions with a standard deviations one-tenth those of the priors;
\item eccentricities are generated uniformly between 0 and 0.01
\item relative longitudes of periastron and true anomalies are generated uniformly between 0 and 360$\degr$.
\end{itemize}
Some of the above parameters are therefore started from a small region of parameter space, before spreading out to explore the prior.
This allows us to judge the burn-in phase to be complete once the trace plot stabilises (see Fig.~\ref{fig:walkers}); we use a burn-in of 2\,000 steps.

\section{Full results for YZ Ceti}
\label{app:fullresults}

\begin{figure*}
\includegraphics[height=0.8\textheight]{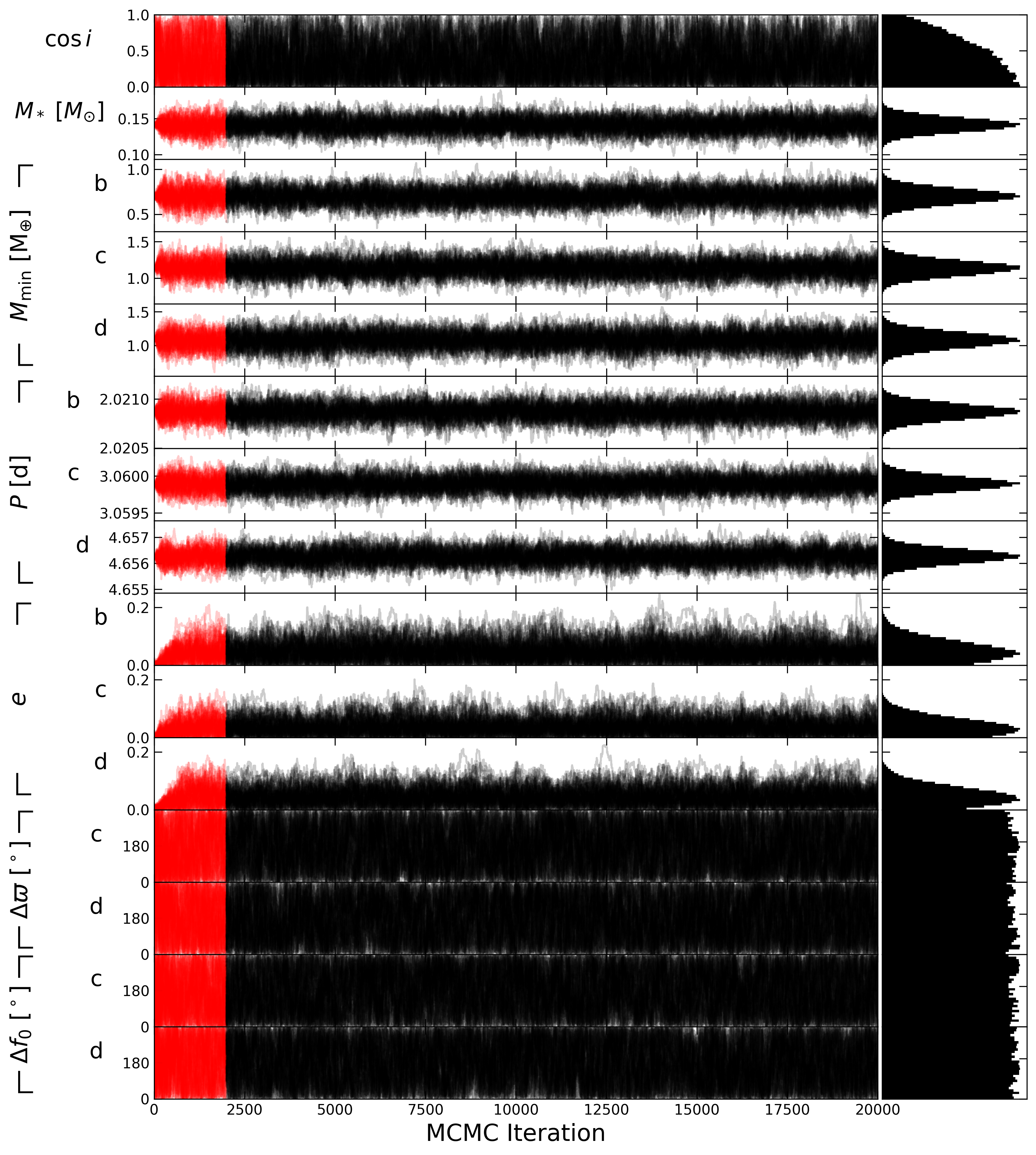}
\centering
\caption{
    Walkers from an MCMC run applying \cinemas{} to the YZ Ceti planetary system.
    The convergence of the eccentricities (rows 9--11) to the posterior show that a burn-in of 2000 links is appropriate for this system.
    The inclination (top row) is parametrised as the prior-uniform $\cos i$, and converges to a distribution which skews towards larger angles compared to the prior (Fig.~\ref{fig:1dposteriors}).
    The longitudes of periastron and initial true anomalies are not constrained (bottom four rows).
}
\label{fig:walkers}
\end{figure*}

The trace plot for an MCMC run applying \cinemas{} to the YZ Ceti planetary system (Fig.~\ref{fig:walkers}) shows that the walkers converge to the posterior distribution adequately within 2\,000 links.
A similar rate of convergence is found for the other systems studied in this work.

\begin{figure*}
\centering
\includegraphics[width=\textwidth]{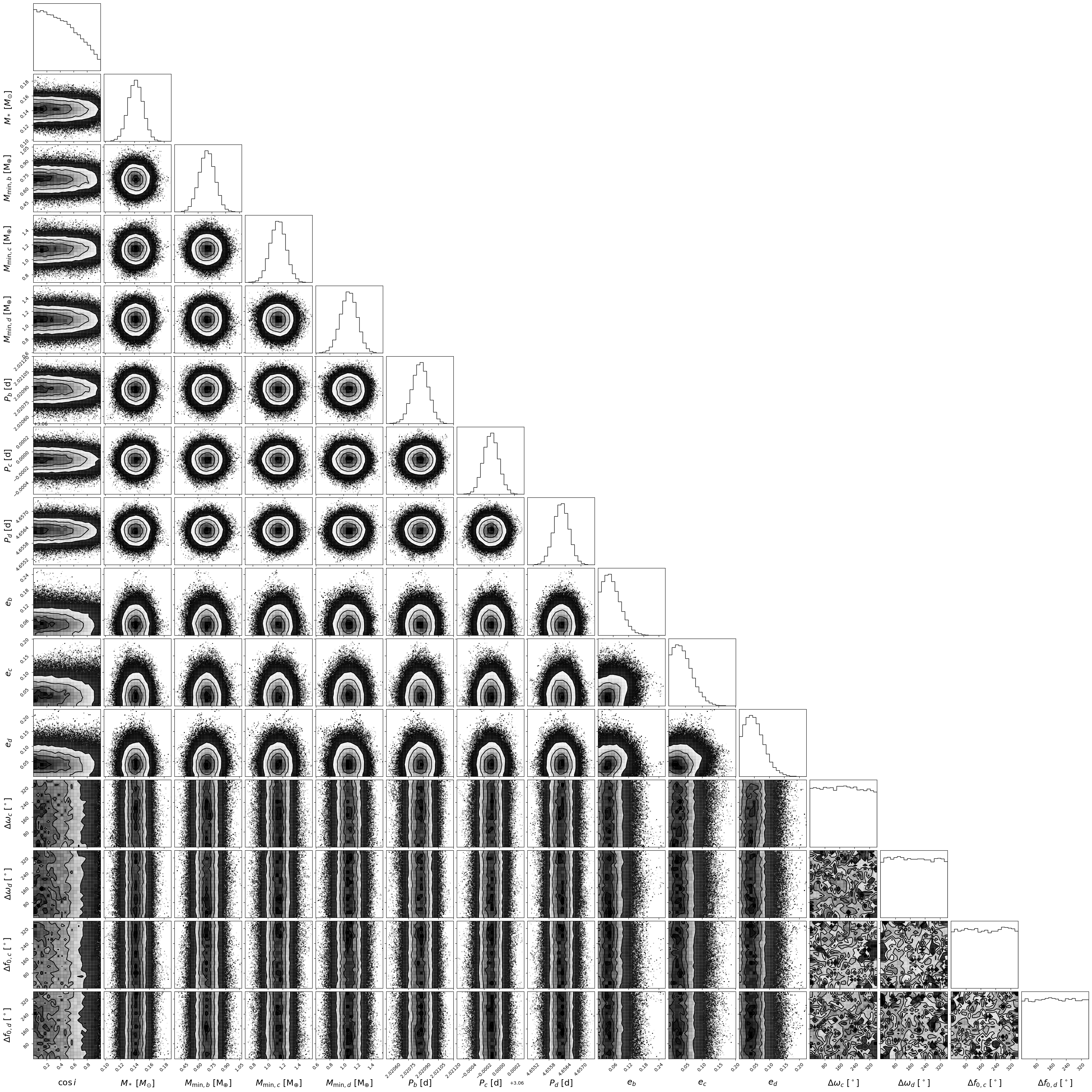}
\caption{
    Corner plot showing the posterior on the parameters of the YZ Ceti planetary system.
    There is no clear covariance between any pair of parameters.
}
\label{fig:corner}
\end{figure*}

Fig.~\ref{fig:corner} shows the corner plot for the \cinemas{} analysis of the YZ Ceti planetary system.
The posterior shows no clear covariance between any of the parameters.

%%%%%%%%%%%%%%%%%%%%%%%%%%%%%%%%%%%%%%%%%%%%%%%%%%

% Don't change these lines
\bsp	% typesetting comment
\label{lastpage}
\end{document}